\documentclass[12pt]{article}
\usepackage{times,color}
\usepackage{amsmath}
\usepackage{amssymb}
\usepackage{xcolor} 
\usepackage{subfigure,epsfig,graphicx,epstopdf}
\usepackage[colorlinks]{hyperref}
\hypersetup{linkcolor = {black}, citecolor = {black}, urlcolor = {black}}
\usepackage[
    backend=biber,
    style=nature,
    sorting=none
]{biblatex}
\AtEveryBibitem{\clearfield{language}}
\AtEveryBibitem{\clearfield{issn}}
\AtEveryCitekey{\clearfield{issn}}
\AtEveryBibitem{\clearfield{url}}
\AtEveryBibitem{\clearfield{month}}
\AtEveryBibitem{\clearfield{urldate}}
\AtEveryBibitem{\clearfield{date}}

\usepackage{upgreek}

\usepackage{caption}
\DeclareCaptionLabelFormat{bflabel}{\textbf{#1~#2}} 
\newenvironment{sciabstract}{%
\begin{quote} \bf}
{\end{quote}}

\newcounter{lastnote}

\begin{document}

\title{A Chip-scale Space-time Multiplexed Gaussian Boson Sampling Processor Beyond 10,000 Photons}
\author{Yu-Xuan Fu$^{1}$, He-Yu Shen$^{1}$, Ke-Ming Hu$^{2}$, Jun-Jie He$^{2}$, Yun-Long Nie$^{1}$, \\
Hang Song$^{1}$, Bao-Jing Liu$^{1}$, Le-Si Yang$^{2}$, Xiao-Yu Wu$^{2}$, Pei-Lin Du$^{2}$, \\
Yu-Ze Zhu$^{2}$, Yi Xie$^{1}$, De-Hui Huang$^{1}$, Yu-Fei Liu$^{2}$, Hai Yan$^{2}$, Jin-Hong Chen$^{3}$, \\
Yu-Lin Yu$^{2}$, Chuan-Yan Peng$^{2}$, Wen-Hao Zhou$^{1}$, Feng Lu$^{1}$, \\
Yu-Quan Peng$^{2}$, Chen-Shuo Xia$^{2}$, Zhi-Chao Wang$^{2}$, Zhe-Han Li$^{2}$, Lin Chen$^{2}$, \\
Chao-Yang Zhang$^{2}$, Chun-Yan Jin$^{2}$, Yan-Rong Zong$^{2}$, Xiao-Yun Xu$^{1}$, \\
Jian-Peng Dou$^{4}$, Xiao-Tian Fang$^{4}$, Pei-Qi Zhou$^{2}$, Hao Tang$^{1,\ast}$, \\
Chang-Shun Wang$^{1,\ast}$, Lin Yang$^{2,\dag}$, Xian-Min Jin$^{1,2,3,4,5,\ddag}$ \\
\normalsize{$^1$Center for Integrated Quantum Information Technologies (IQIT), School of Physics}\\
\normalsize{and Astronomy and State Key Laboratory of Photonics and Communications,}\\
\normalsize{Shanghai Jiao Tong University, Shanghai 200240, China}\\
\normalsize{$^2$TuringQ Co., Ltd., Shanghai 200240, China}\\
\normalsize{$^3$Chip Hub for Integrated Photonics Xplore (CHIPX),}\\
\normalsize{Shanghai Jiao Tong University, Wuxi 214000, China}\\
\normalsize{$^4$Atomology Co., Ltd., Shanghai 200240, China}\\
\normalsize{$^5$Hefei National Laboratory, Hefei 230088, China}\\
\normalsize{$\ast $E-mail: htang2015@sjtu.edu.cn}\\
\normalsize{$\ast $E-mail: cswang@sjtu.edu.cn}\\
\normalsize{$\dag$E-mail: yanglin@turingq.com}\\
\normalsize{$\ddag$E-mail: xianmin.jin@sjtu.edu.cn}
}

\date{}
\baselineskip24pt

\maketitle

\begin{sciabstract}
Gaussian boson sampling (GBS) has emerged as a leading photonic paradigm for demonstrating quantum computational advantage.
Nevertheless, state-of-the-art GBS setups face practical barriers including stringent optical alignment, phase instability, and limited programmability, which impede scalable engineering deployment.
The chip-scale space-time multiplexed architecture promises to resolve these constraints, yet it strongly demands wafer-scale chip capabilities to simultaneously satisfy stringent requirements on low loss, high precision and high-speed modulation.
Here we report the first chip-scale space-time multiplexed GBS system, monolithically integrating high-speed electro-optic modulators, on-chip delay lines, and a time-space multiplexed interferometric network on a thin-film lithium niobate chip, operating at a 4-GHz clock rate with detection events of up to 11,059 photons within 1 millisecond.
Beyond benchmarking quantum advantage, we further reconfigure the photonic hardware into a GBS-powered world model for modelling physical dynamics, which achieves lower prediction error with fewer trainable readout parameters compared with a classical echo state network (ESN) baseline.
Our results validate the feasibility of our endeavor towards scalable photonic quantum hardware, and pave the way for the versatile programmable applications of future GBS quantum systems.
\\
\end{sciabstract}

\section*{Introduction}
\noindent 
The pursuit of quantum computational advantage, outperforming the state-of-the-art classical supercomputers on targeted tasks, stands as one of the most pivotal milestones in quantum computing. Since the superconducting Sycamore processor first claimed such an advantage~\cite{Arute2019}, photonic systems have emerged and these remain the two physical quantum systems that can achieve the advantage up to now~\cite{Wu2021PRL, Liu2026}, carving out their own distinct routes to scalability.
In photonic systems, boson sampling~\cite{Aaronson2011,Spring2013,Broome2013,Tillmann2013,Crespi2013}, together with its variants, particularly Gaussian boson sampling (GBS)~\cite{Hamilton2017}, is a representative paradigm for demonstrating quantum advantage, as exactly evaluating its output probabilities is \#P-hard in the general case.
A batch of experiments has demonstrated quantum computational advantage in photonics~\cite{Gao2022, Madsen2022,Zhong2020, Zhong2021, Deng2023, Liu2026}. While some among them established the free-space route through cascaded interferometers, Zhiyuan~\cite{Gao2022} and Borealis~\cite{Madsen2022} introduced time-domain multiplexed schemes and demonstrated the realization of a large-scale equivalent interferometric network with modest physical resources.
In recent years, the classical simulation capacity keeps increasing, and high loss has been shown to undermine the classical intractability of GBS~\cite{Oh2024}, which raises more stringent requirements for claiming quantum computational advantage.  

With growing system scale and complexity, stability and manufacturability emerge as the primary practical bottlenecks to engineering deployment. The proof-of-principle implementations in free-space bulk optics greatly suffer from the difficulties in meticulous alignment, the lack of stability, and the extremely weak programmability, which is detrimental for application research. All-Fiber-based approaches, while having confirmed the feasibility of time-domain multiplexing strategy~\cite{Humphreys2013,Madsen2022,Motes2014,He2017,Asavanant2019,Larsen2019,Yonezu2023, Monika2025,Borghi2025,Basani2026,SempereLlagostera2022,Yu2023}, may still be limited by the phase stability and the lack of utilization of spatial resources.

We provide a holistic assessment of the challenges and propose that the route has to resort to chip-scale integration: when interleaving multi-loop temporal cascades and multi-layer Mach-Zehnder interferometer (MZI) spatial networks on chip, complemented by thermo-optic and high-speed electro-optic phase shifters~\cite{Shang2026,Nie2026}, full mixing of temporal and spatial modes can then be precisely engineered on demand, to fundamentally address the issue of scalability and programmability. In order to enable the space-time multiplexing scheme~\cite{He2025Chip}, we have to realize extremely high modulation speed to allow for moderate delay loop length on chip for time-bin operations, and we have to keep lowering down various losses on chip to ensure spatial scalability. Only when simultaneously satisfying all these engineering challenges can we surpass the bottlenecks for space-time multiplexing on chip.
 
In this paper, we report the first chip-scale space-time multiplexed GBS system, Zhiyuan 3.0, with the largest chip-scale GBS mode number and detection events up to 11,059 photons within 1 millisecond. With endeavors on setting up wafer-scale thin-film lithium niobate (TFLN) pilot line and iterative optimization within a closed-loop fabrication workflow, we successfully integrate high-speed electro-optic modulators, on-chip waveguide delay lines, and the time-space multiplexed interferometer network on a single chip, reaching a system clock rate of 4 GHz.
Beyond demonstrating the quantum computational advantage, we explore the essential link between the intrinsic sampling of high-dimensional mode correlations in GBS and the data-driven artificial intelligence (AI) research exemplified by the world model. Taking the K\'arm\'an vortex street as a representative physical scenario, we reprogram the same chip into a GBS-powered world model that learns and predicts the flow-field dynamics.
Compared with a classical echo state network (ESN) baseline, the model achieves lower prediction error with fewer trainable readout parameters, demonstrating the potential of photonic quantum hardware for AI.

\section*{Results}
\begin{figure*}
\centering
\includegraphics[width=1 \columnwidth]{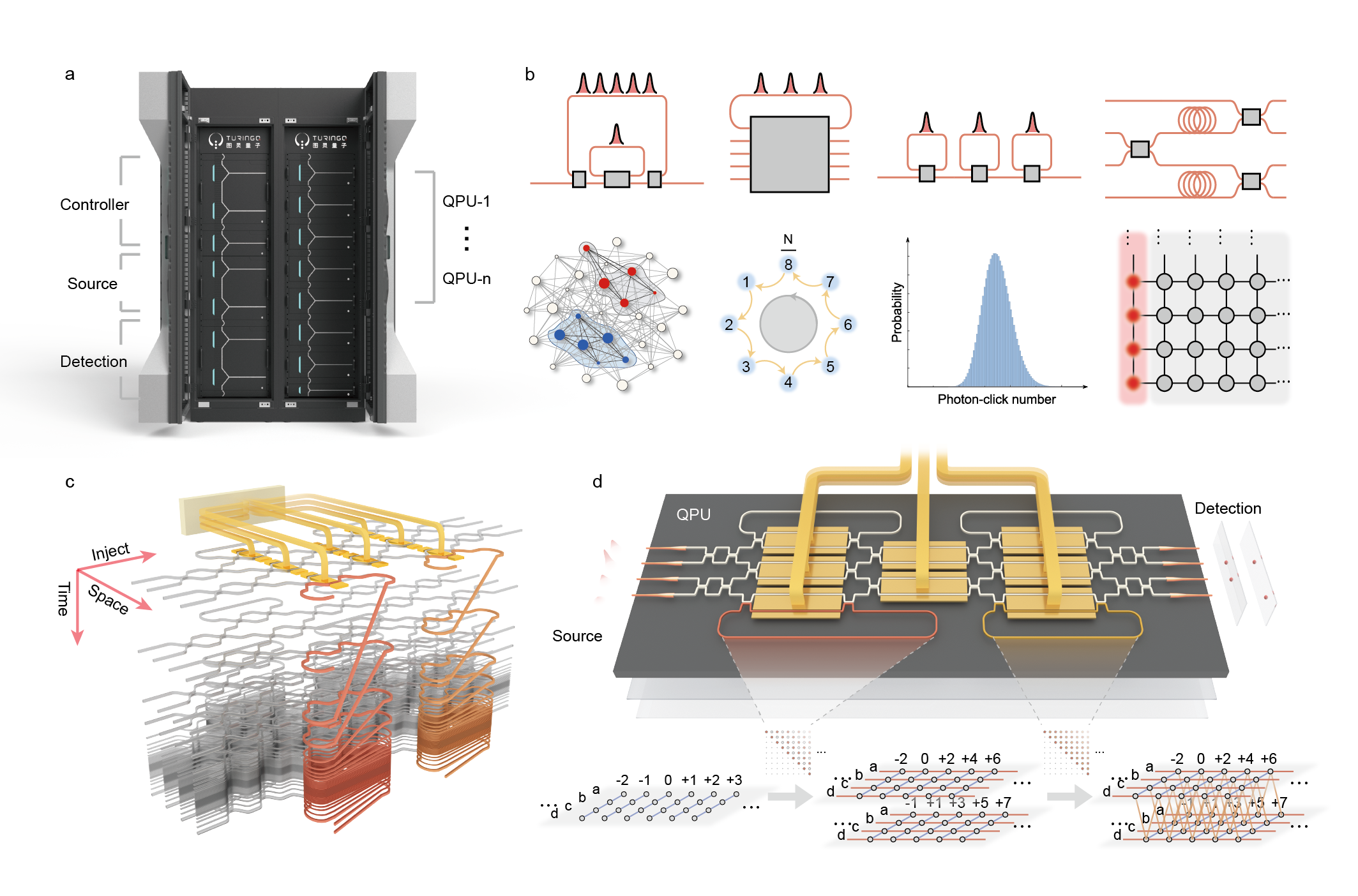}
\caption{\textbf{Schematic diagram of the chip-scale space-time multiplexed GBS system Zhiyuan 3.0.} \textbf{a} The complete system is comprised of the control module, quantum source, quantum processing unit (QPU) and the detection module. \textbf{b} Top: representative time-domain multiplexing proposals spanning delay-loop and unbalanced delay-line architectures. The four panels, from left to right, show the nested delay-loop firstly proposed in ~\cite{Humphreys2013}, loop-back~\cite{Gao2022}, cascaded delay-loop~\cite{Madsen2022}, and unbalanced delay-line structures~\cite{Asavanant2019,Larsen2019}, respectively. The time-domain multiplexing idea was first proposed in~\cite{Humphreys2013}. Bottom: GBS-related problems among quantum advantage demonstrations, graph problems and cluster-state preparation. \textbf{c} Space-time multiplexing chip architecture. The chip monolithically integrates a multi-layer MZI network with thermo-optic, electro-optic modulator arrays and on-chip delay loops, constructing a large-scale, fully space-time-mixed and highly programmable network. \textbf{d} Squeezed vacuum is launched into the chip and evolves through the cascaded space-time interference structure, giving rise to a space-time-mixed, large-scale GBS state. The state is detected by the superconducting nanowire single-photon detectors (SNSPDs) for sampling, where all spatial modes are detected via pseudo-photon-number-resolving detection~\cite{Cheng2023,Eaton2023,bressanini2024gaussian} using a total of 16 channels.}
\label{fig1}
\end{figure*}

\noindent 
The entire GBS system is housed in a single rack (Fig.~\ref{fig1}a), which contains the modules of the photon source, quantum processing unit (QPU) on chip, photon detection and the control system. As shown in Fig.~\ref{fig1}b, the flexibility of space-time multiplexing has been widely demonstrated in recent years, with delay-loop and unbalanced delay-line proposals spanning quantum advantage, graph problems, and cluster-state preparation. Here we implement this space-time multiplexing strategy onto an integrated photonic chip, empowering strong programmability while preserving the scalability of this scheme.

As shown in Fig.~\ref{fig1}c, the interferometer network is constructed across both space and time. In each layer, the multi-layer MZI networks allow the photon to evolve across different spatial modes. The two delay loops of different lengths connect photons of different time bins. Such an MZI network with delay loops, assisted by thermo-optic and electro-optic modulation capabilities, endows the system with flexible programmability in both space and time dimensions, enabling a single chip to be reconfigured for different tasks. As depicted in Fig.~\ref{fig1}d, a squeezed vacuum state is injected into the chip and the pulses are routed through a sequence of MZIs and delay loops to generate a large-scale three-dimensional cluster state with space-time entanglement. The state is finally delivered to the superconducting nanowire single-photon detectors (SNSPDs) for sampling.

The chip is fabricated by wafer-scale processing on a six-inch thin-film lithium niobate wafer. TFLN has emerged as a vital research platform in recent years owing to its superior electro-optic properties. Various high-performance optoelectronic modulators~\cite{Yu2022,Stokowski2024,Li2023} demonstrating ultra-high-speed light manipulation capabilities~\cite{He2019,Valdez2023}, as well as multiple frequency-conversion devices~\cite{Yuan2021, Wu2024, Fang2025,Shi2026}, have been realized on this material system. While TFLN modulators have yielded preliminary quantum-information demonstrations~\cite{Labbe2025,Zhang2026,Sund2023}, the full power of their large-scale high-speed modulation remains not truly harnessed for quantum-computing applications.

\begin{figure*}
\centering
\includegraphics[width=1 \columnwidth]{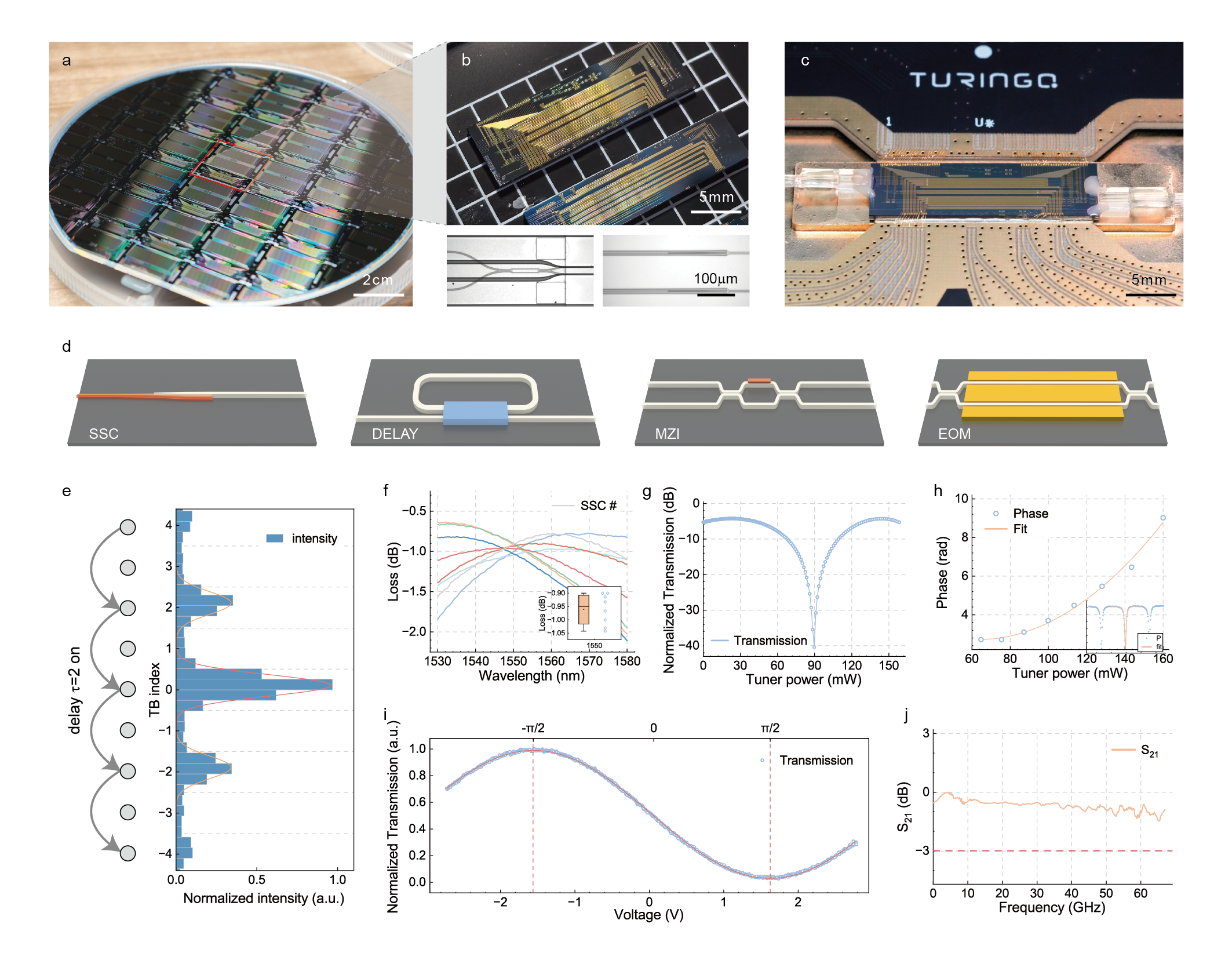}
\caption{\textbf{Chip fabrication and characterization.} \textbf{a-c} Layout of the lithium niobate-based TDM-GBS chip. Thin-film lithium niobate wafer (a), fabricated chip (b), and packaged system module (c).  \textbf{d} Illustrations of the key devices: dual-layer spot-size converter (SSC), delay loop, thermo-optic MZI, and the high-speed electro-optic modulator (EOM). \textbf{e-j} Device characterization. \textbf{e} The system operates at a 4-GHz clock rate; the 250-ps time-bin windows can be clearly divided on the time axis. The red line shows the fit of the time-of-arrival and the gray dashed lines mark the time-bin windows. \textbf{f} Transmission spectra of multiple SSCs measured over 1530-1580 nm. The inset shows the loss distribution at 1550 nm, with a mean of 0.9 dB per facet. \textbf{g} MZI transmission versus thermo-optic tuning power with extinction ratio above 35 dB. \textbf{h} Modulation curve of the delay loop, characterized by the shift of the resonance peak. \textbf{i} Half-wave voltage of the EOM, $V_{\pi} = 3.18 V$. \textbf{j} Measured 3-dB bandwidths of on-chip modulators.}
\label{fig2}
\end{figure*}

Recall that the loss budget imposed by the criterion~\cite{Oh2024} is extremely stringent and technically challenging to satisfy. To meet this rigorous loss constraint, we have dedicated 4.5 years to establishing a complete wafer-scale lithium niobate photonic chip pilot line. Throughout the development, every structural and functional element was repeatedly optimized and verified via closed-loop fabrication iterations, with each full iteration cycle compressed to three weeks. Through exhaustive parameter tuning and laborious iterative refinement, we ultimately fulfilled the requirements defined by the criterion~\cite{Oh2024}. We fabricate the chip (see Methods) with the low-loss interferometer network, high-speed modulator arrays, and delay loops integrated within a single shot. We then package it via fiber-array attachment and wire bonding, and finally assemble it into the experimental system, as shown in Fig.~\ref{fig2}a-c. The detailed structures for key devices are shown in Fig.~\ref{fig2}d.

To evaluate the chip-scale performance achievable with the current fabrication process, we perform device characterization on the wafer-scale run, as shown in Fig.~\ref{fig2}e-j. The system operates at a 4-GHz clock rate, the highest reported to date in quantum systems, with 250-ps time-bin windows. Fig.~\ref{fig2}e shows an example of the coincidence time-difference distribution under single-pulse injection, showing that the time-bin windows can be clearly divided on the time axis. The spot-size converters exhibit uniform transmission over 1530-1580 nm, with a mean loss of 0.9 dB per facet at 1550 nm. The thermo-optic MZIs exhibit clear interference fringes in the transmission spectra as the heating power varies, with an extinction ratio above 35 dB, as depicted in Fig.~\ref{fig2}g. In Fig.~\ref{fig2}h, phase modulation on the delay lines is characterized via the shift of the resonance peak. The electro-optic modulators achieve a half-wave voltage of 3.18 V, and the electro-optic S21 response yields a 3-dB bandwidth exceeding 67 GHz, as shown in Fig.~\ref{fig2}i-j, supporting the system operating at the 4-GHz clock rate. The entire system is temperature-stabilized to within 0.01 $^{\circ}\mathrm{C}$.

\begin{figure*}
\centering
\includegraphics[width=1 \columnwidth]{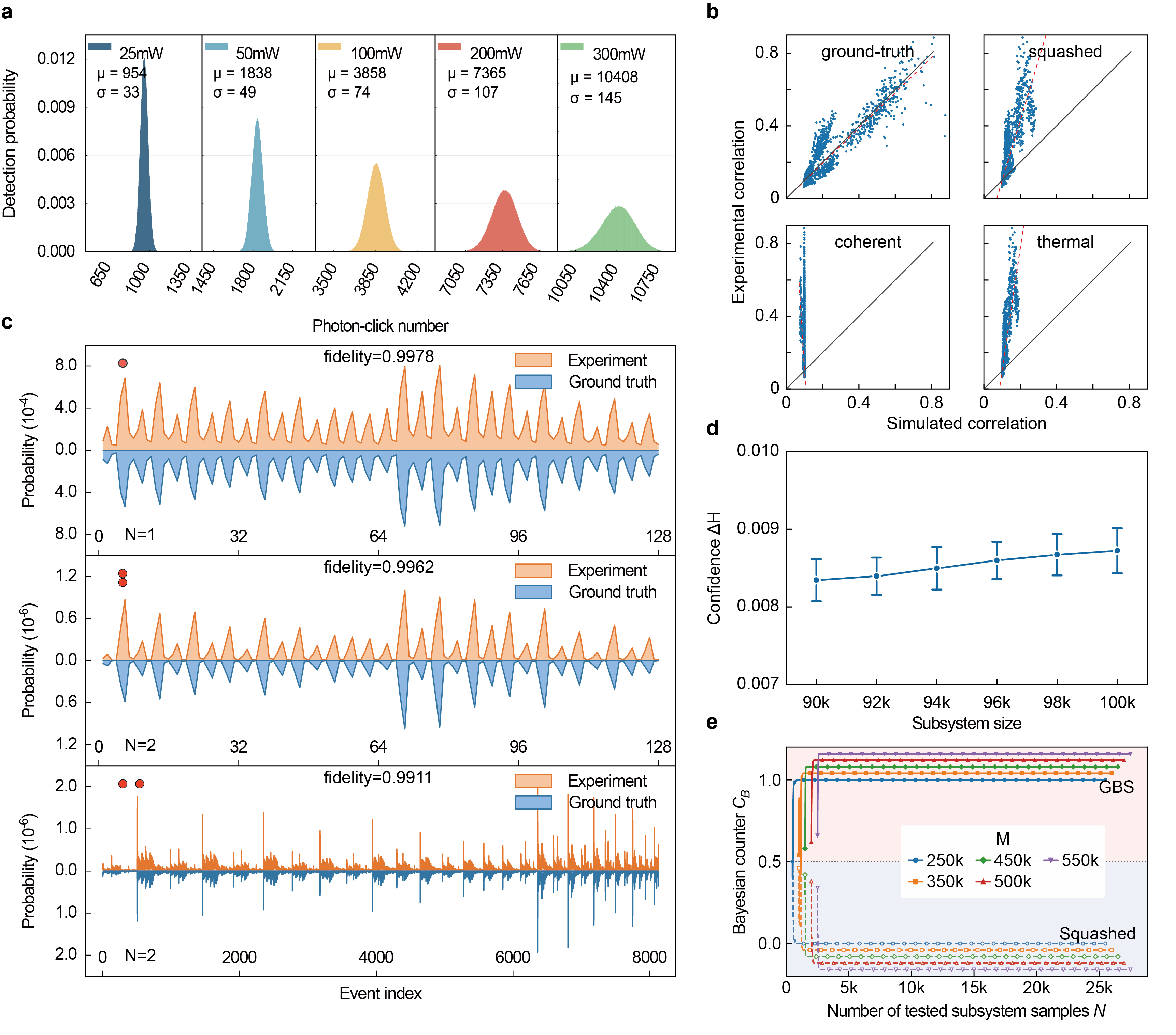}
\caption{\textbf{Experimental results and validation of the GBS device.} \textbf{a} Normalized probability distributions of the photon-click number for five pump powers. At the highest pump power, the largest photon-click number exceeds 10,000. \textbf{b-c} Small-scale GBS validation. \textbf{b} Simulation-experiment comparison of second-order correlation functions under four hypotheses with the same first-order moments, confirming that the photon statistics of the experimental output agree with the GBS ground truth. \textbf{c} Per-configuration comparison between experimental and theoretical probabilities in the few-photon subspace, with single photon, two-photon bunching, and two-photon non-bunching fidelities $0.9978$, $0.9962$, and $0.9911$. \textbf{d} Bayesian confidence. All scores are above zero, indicating that the samples generated by this system lie closer to the ground-truth distribution than to any adversarial spoofer distribution. \textbf{e} Bayesian counter. Batch-wise evolution of the score, with the discrimination between the GBS ground truth and each spoofer steadily diverging as the number of samples increases, confirming the GBS model.}
\label{fig3}
\end{figure*}

We conduct the sampling experiments using five different pump powers and obtain the normalized probability distributions of the total click number per sample, each corresponding to a sampling process within 1 millisecond. We show the distributions shifting towards larger photon numbers as the pump power increases. As depicted in Fig.~\ref{fig3}a, at the highest pump power of 300 mW, the largest click number in a single sample was up to 11,059---a value well beyond any previously reported in GBS experiments, rendering precise reproduction of the output distribution beyond classical computational capacity (see Methods for more details).

To verify the statistical properties of the experimental output, we first carry out validation on a small-scale system. As shown in  Fig.~\ref{fig3}b, we compute the photon correlation function theoretically under four hypotheses—the GBS ground truth, thermal light, coherent light, and the squashed state—and compare them with the experimental results. The first-order moments (mean photon number per mode) are set according to the experimentally calibrated values across all hypotheses for a fair comparison. Only the GBS hypothesis yields data points that lie along the diagonal, whereas the other three models exhibit clear deviations, confirming that the output photon statistics are consistent with the GBS theory. Besides, we compare the theoretical and experimental probability distributions for each output configuration of the few-photon coincidences. The fidelities between the theoretical and experimental results for the three cases—single photon, two-photon bunching, and two-photon non-bunching—are $0.9978$, $0.9962$, and $0.9911$, respectively, as shown in  Fig.~\ref{fig3}c, validating the experimental distributions.

We further perform Bayesian analysis~\cite{Qi2022} on the experimental data to validate the system's output statistics at the model level (see Methods for more details). First, the Bayesian confidence check is applied to a 30-photon subsystem and all confidence scores are positive and increase with the subsystem size, suggesting that the generated samples are closer to the true GBS distribution than to the spoofer, as shown in Fig.~\ref{fig3}d. Besides, taking the squashed spoofer as an example, the Bayesian counter traces the evolution of the coincidence score as samples accumulate in batches, showing that the GBS hypothesis becomes progressively more distinguishable from the squashed spoofer, as depicted in Fig.~\ref{fig3}e, confirming that the output is genuine GBS rather than any classical spoofing model. Together with the precise few-photon distribution comparisons, the Bayesian analysis establishes that the system's output matches GBS at both the statistical and the model level.

Sampling at the scale of thousands of photons and the demonstration of quantum advantage address only the question of ``how large can it be''. In recent years, research on GBS—including the results of this work—has been driven by a sustained race for scale, with mode numbers and photon numbers rising steadily over time. Yet claiming quantum advantage is a necessary milestone, not the final destination. The next step of this work is to pursue a practical aspect, which calls for three engineering qualities: economy through wafer-scale mass fabrication, robustness through on-chip integration, and usability through flexible reconfiguration. The combination of chip-scale integration, space-time mixing, and high-speed programmability transforms this system from a merely large GBS into a usable one, as demonstrated by the GBS-powered photonic-native world model below.

\begin{figure*}
\centering
\includegraphics[width=1\columnwidth]{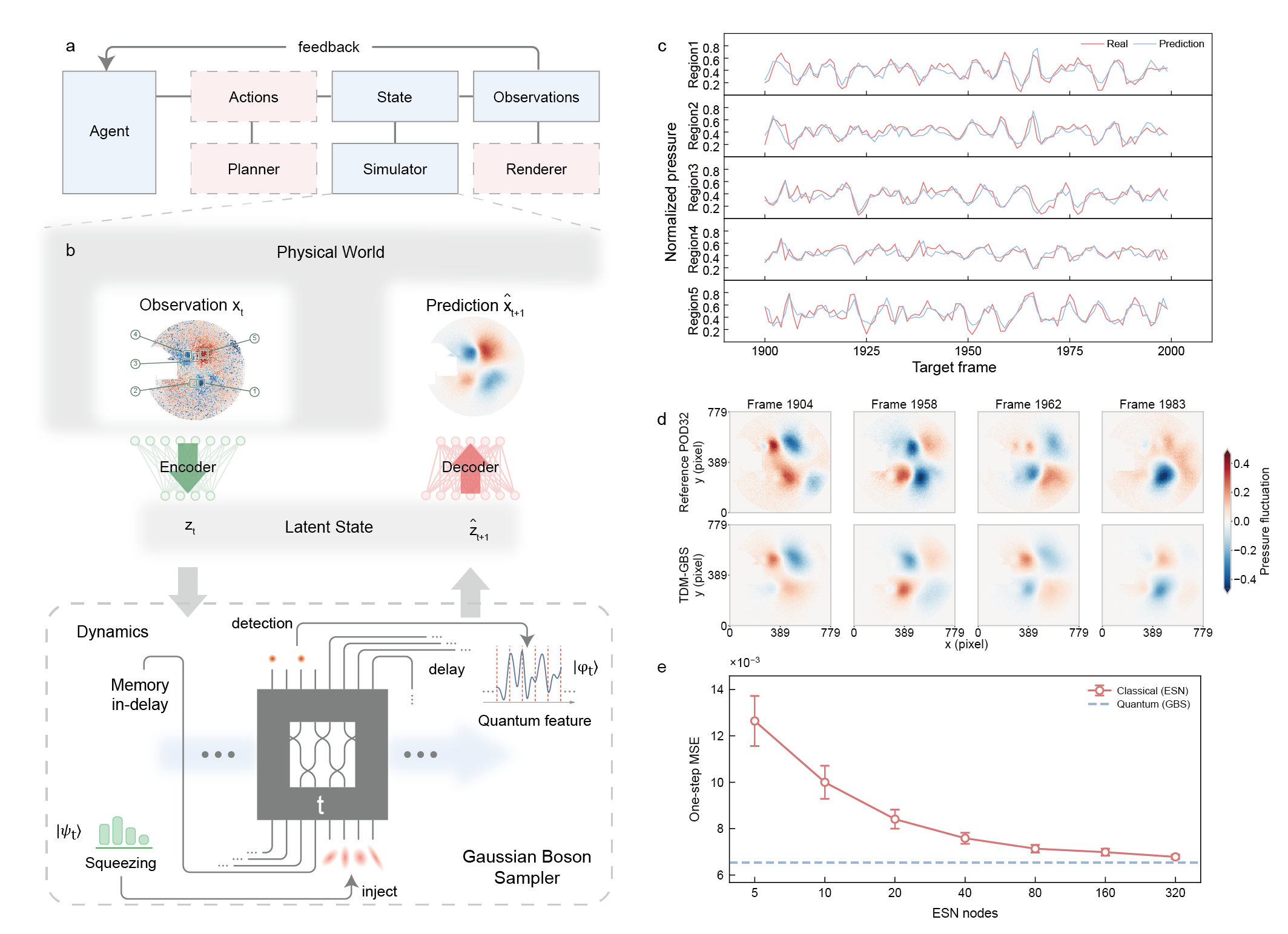}
\caption{\textbf{GBS-powered world model.} \textbf{a} Functional taxonomy of world models--Planner, Simulator, and Renderer--with our system falling into the Simulator category. \textbf{b} Closed-loop framework of the GBS-powered world model: Physical World $\to$ Observation $\to$ Encoder $\to$ GBS Dynamics model and Memory $\to$ Decoder $\to$ future-state Prediction, with GBS serving as the dynamics engine in the loop. \textbf{c} Comparison between the latent-variable evolution of the true vortex street (red) and the GBS prediction (blue). \textbf{d} Representative flow-field reconstruction: the top row shows the original vortex-street flow field and the bottom row the flow field reconstructed from GBS-powered world model predictions. \textbf{e} Prediction errors of the GBS-powered world model versus the classical method (ESN) on the K\'arm\'an dataset.}
\label{fig4}
\end{figure*}

A growing body of research is devoted to the synergy of quantum computing and AI, where quantum states and programmable dynamics serve as native resources for data-driven learning.~\cite{Sharifian2026, Cimini2026}.
World models give this convergence a predictive focus: learning representations of physical dynamics from observations and using them to forecast future states~\cite{Ha2018, Hafner2023}.
These predictive representations support applications in planning, simulation and rendering (Fig.~\ref{fig4}a).
Recent work on quantum-enhanced long short-term memory networks and quantum reservoir computing has demonstrated temporal prediction using quantum dynamical systems~\cite{Chen2020,Takagi2025,Hou2026,Paparelle2026}.
Motivated by these advances, we implement a GBS-powered world model with the reconfigured chip as its dynamics engine.
This photonic quantum world model exploits photonic-native temporal memory: optical delay loops retain part of the evolving optical state and mix it with subsequent inputs.
Interference and photon counting then yield features that encode information about the input history.

We demonstrate the model through one-step prediction of K\'arm\'an vortex-street pressure dynamics~\cite{Takagi2025}.
In the architecture shown in Fig.~\ref{fig4}b, the encoder maps five regional means of the observed pressure field to optical inputs.
A trained linear readout uses the measured photonic features to predict the next five-dimensional pressure state, which a classical spatial decoder maps back to a two-dimensional field (see Methods).
Each prediction uses the observed current state together with the internal optical memory.
Fig.~\ref{fig4}c compares the reference and predicted regional-mean pressures (red: reference; blue: prediction).
Fig.~\ref{fig4}d compares the predicted pressure-field reconstructions with the rank-32 proper orthogonal decomposition (POD) reconstructions of the reference fields at representative frames, showing agreement in the main spatial structures.

We benchmark the GBS-powered world model against classical ESNs using the same requested squeezing inputs and temporal evaluation segment (Fig.~\ref{fig4}e).
Although the ESN prediction error decreases as the network size increases from 5 to 320 nodes, the photonic quantum world model maintains a lower mean squared error (MSE) than the ESN mean at every tested size on the final 100 frames.
The experimentally measured photonic features yield a one-step MSE of $6.537\times10^{-3}$, compared with $(6.781\pm0.091)\times10^{-3}$ for the 320-node ESN, where the uncertainty denotes the sample standard deviation across 100 random initializations.
The quantum model thus reduces the MSE by $3.60\%$ relative to the mean MSE of the largest evaluated ESN.
Its error also lies below the lowest error among the 100 evaluated ESN initializations at each tested network size on this segment.

The photonic model achieves this accuracy with $84.1\%$ fewer trainable readout parameters than the 320-node ESN (see Methods).
Together, the lower prediction error and smaller trainable readout demonstrate the potential of the GBS-powered world model for parameter-efficient dynamical prediction on the recorded trajectory, extending the use of reconfigurable photonic quantum hardware to AI tasks beyond sampling benchmarks.

\section*{Discussion}
\noindent 

In conclusion, we demonstrate the first space-time multiplexed GBS in the chip scale, with the interferometer network, modulator arrays, and delay loops monolithically integrated on a thin-film lithium niobate platform. The system operates at a 4-GHz clock rate and produces samples with more than 10000 photons at the largest scale. We validate the results with photon correlation functions, few-photon probability distributions, and Bayesian analysis, collectively confirming that the system output is consistent with GBS ground truth. As for the complexity, we evaluate the classical simulation cost using the matrix product state (MPS) method, the best available classical algorithm, showing that classical simulation at this scale is computationally prohibitive. The GBS-based world model built on the same chip further demonstrates the system's capability for practical computational tasks beyond quantum-advantage demonstrations. Overall, these results demonstrate that our chip-scale space-time multiplexed system offers both scale and programmability, evolving GBS from a dedicated sampling experiment into a versatile quantum computing platform.

As discussed above, the race for scale is not an end in itself; the next direction is to move from scale competition toward practical applications. The engineering qualities of the system—economy, robustness, and usability—serve as the benchmarks for future progress. Therefore, it will be necessary to integrate squeezed light sources and detectors on chip to eliminate the remaining free-space components, while lowering waveguide and coupling losses to further extend the system scale~\cite{PsiQuantumTeam2025}. The co-packaged optics techniques~\cite{Kim2026,Hua2025,Ahmed2025}, as well as high-speed electrical readout and control systems~\cite{Caldwell2025,Bazammul2025,Duggan2026}, are also potentially great benefits for scalable photonic quantum processing units. From a broader perspective, the significance of chip-scale GBS extends beyond quantum-advantage demonstration. The GBS-powered world model exemplifies the convergence of quantum hardware and AI by coupling photonic-native temporal memory to learned predictions of physical dynamics. As a programmable photonic quantum platform, it provides a hardware foundation for various photonic quantum computing tasks, with an eventual goal of fault-tolerant photonic quantum computation~\cite{AghaeeRad2025,Konno2024,Larsen2025Nature,Yu2026NatPhoton}. Fully leveraging quantum engineering thus empowers chip-scale photonic quantum computing platforms and broadens the horizons for practical explorations.

\section*{Methods}
\subsection*{Wafer-scale fabrication of low-loss, high-speed chips}  

The chip is fabricated on 6-inch commercially available thin-film lithium niobate on insulator wafers, which consist of a stack with a 400 nm thick mono-crystal x-cut LiNbO3 layer on top of a 4.7 $\mu m$ buried thermal oxide (BOX) layer (NanoLN). Waveguides are patterned using a deep-ultraviolet (DUV) scanner (KrF) and then transferred to the lithium niobate (LN) layer by etching 200 nm of LN via an Ar$^{+}$-based etching process. The waveguide layer is protected by a 1200 nm silicon oxide cladding deposited by plasma-enhanced chemical vapor deposition. Heaters made of titanium nitride with 175 nm thickness are fabricated using an etching process. Metal electrodes made of gold are fabricated using a lift-off process with a 1000 nm thickness for the unloaded electrodes. Finally, the silicon substrate under the heater section is removed by isotropic plasma silicon dry etching.
 
\subsection*{GBS validation metrics}
\noindent \textbf{Few-photon distributions}

To quantitatively assess the accuracy of the experimental results, we accumulated few-photon event statistics over $32$ time steps and computed the fidelities of the single-photon, the two-photon bunching and non-bunching events separately. With $p_i$ and $q_i$ denoting the experimental and theoretical distributions, the fidelity (Bhattacharyya coefficient) is defined as
\begin{equation}
F=\sum_i \sqrt{p_i\,q_i}\,.
\end{equation}
The $32$ time steps comprise $128$ modes in total, yielding $128$, $128$, and $8128$ possible patterns for $P_{1}$, $P_{2}$, and $P_{1,1}$ respectively. In the theoretical simulation, we constructed the space-time multiplexed circuit within the DeepQuantum simulation framework~\cite{He2025DeepQuantum} and unfolded the circuit with 32 time steps into its global equivalent form, thereby efficiently obtaining the corresponding theoretical distributions.

\noindent \textbf{Photon correlation function}

The $g^{(2)}$ matrix is defined through normal ordering as
$
g^{(2)}_{ij}=\frac{\langle :\hat n_i\,\hat n_j:\rangle}
{\langle \hat n_i\rangle\langle \hat n_j\rangle}
$,
with
$
 \hat n_i=\hat{a}_i^\dagger \hat{a}_i
$,
where the diagonal and off-diagonal elements can be further written as
\begin{equation}
g_{ij}^{(2)} = 
\begin{cases} 
\frac{\langle \hat{n}_i (\hat{n}_i - 1) \rangle}{\langle \hat{n}_i \rangle^2}, & i = j \\ 
\frac{\langle \hat{n}_i \hat{n}_j \rangle}{\langle \hat{n}_i \rangle \langle \hat{n}_j \rangle}, & i \neq j 
\end{cases}
\end{equation}

In our space-time-multiplexed architecture, the mode index is defined as $i = 4\times t + s$, 
where $t$ denotes the time step and $s\in \{1,2,3,4\}$ represents the spatial mode index.

To verify the correlation structure and quantum statistical features of the system output, we carried out a small-scale benchmark with 32 time steps (corresponding to 128 modes in total). The experimentally measured $g^{(2)}$ matrix was compared with theoretical predictions for squeezed, coherent, thermal, and squashed states.

\noindent \textbf{Bayesian counter}

For the \( l \)-th experimental sample, let 
$
Q_l = P_{\mathrm{GBS}}(l), R_l = P_{\mathrm{mock}}(l),
$
where $Q_l$ is the probability assigned by the target GBS model, and $R_l $ is the probability assigned by a given mock model.
The cumulative likelihood ratio is defined as
\begin{equation}
\chi_N = \prod_{l=1}^N \frac{Q_l}{R_l}.
\end{equation}
Assuming equal prior probabilities for the GBS and mock models, the posterior probability that the samples originate from the GBS model is given by
\begin{equation}
C_B(N) = \frac{\chi_N}{1 + \chi_N}.
\end{equation}
Thus, $C_B(N) > 0.5$ favors the GBS model, while $ C_B(N) < 0.5 $ favors the mock model.

\noindent \textbf{Subsystem Bayesian confidence}

Assume two hypotheses: $ H_0 $ is the target GBS distribution, and $ H_1 $ is the spoofer distribution. The Bayes confidence is essentially the average log-likelihood ratio of the samples under the two models:
\begin{equation}
\Delta H = \frac{1}{N} \sum_{i=1}^N \ln \frac{P_{H_0}(S_i \mid n)}{P_{H_1}(S_i \mid n)}.
\end{equation}

Here, the total click number $ n $ is fixed. Therefore, the comparison addresses the following question: given the same total number of clicks, the events are more consistent with either the GBS model or the spoofer model. The decision criterion is: if $ \Delta H > 0 $, the experimental samples favor the GBS model.

\subsection*{Quantum advantage benchmark}

To assess the classical simulability of this work, we perform simulation evaluations based on Refs.~\cite{Oh2024, Liu2026}. Ref.~\cite{Oh2024} shows that the covariance matrix with losses can be decomposed into a quantum part $ V_{p} $ and a classical part $ W $, enabling efficient simulation of the lossy GBS distribution via the MPS method, where the simulation complexity is primarily determined by the quantum part $ V_{p} $. For moderate-scale systems, $ V_{p} $ can be obtained via semidefinite programming (SDP). For large-scale systems ($ m > 2000 $), Ref.~\cite{Liu2026} further provides an analytical construction method to approximately generate $ V_{p} $ and estimate the required bond dimension. Based on the above methods, we conduct a benchmark assessment of the classical simulation complexity for our experiment, aiming to quantify the simulability boundary across different system scales and provide a foundation for discussing quantum advantage.

In the numerical simulations, assuming an on-chip delay-line loss of $2\text{ dB}/\tau$, a fiber delay-line loss of $2\text{ dB}$, a spatial loss of $5\text{ dB}$, and $10,000$ time steps ($M = 40,000\text{ modes}$), the effective photon number $N_{\mathrm{eff}}$ reaches $117.9$. For $d = 3$ and a bond dimension of $\chi \sim 10^4$, the truncation error exceeds $0.05$. According to the computational complexity
\begin{equation}
T_{\mathrm{MPS}} = \mathcal{O}\left(M d \chi^2 2^{N_{\mathrm{eff}}/2}\right),
\end{equation}
and the NVIDIA A100 graphics processing unit (GPU) floating-point operations per second (FLOPS) benchmark, even with sufficient memory, classical simulation of a $2.5\text{-}\mu\mathrm{s}$ GBS sampling task would require at least $9.7 \times 10^6$ years.

\subsection*{GBS-powered world model} 
Dynamical learning seeks to infer physical evolution rules from observational measurements to anticipate subsequent state transitions. Stemming from early cognitive-science insights, researchers have developed physics-aware modeling approaches such as physics-informed neural networks, which embed governing equations as loss constraints to enforce physical consistency. Yet such equation-dependent methods may be challenged by complex, partially known, multimodal and multi-scale real-world scenarios. In contrast, world-model frameworks extract universal, transferable dynamical representations purely from observational data, without complete prior knowledge of underlying physics.

We study the prediction of partially observed K\'arm\'an pressure dynamics on a $780\times780$ grid using 2,000 snapshots, split chronologically into 1,900 for training (1,500 fitting + 400 validation) and 100 for evaluation. The world model consists of an encoder, a dynamical module, and a spatial decoder. The encoder compresses each pressure field into five regional means, which are standardized, clipped, and quantized into 64-level squeezing values. The dynamical module—implemented either by five parallel time-domain-multiplexed GBS (TDM-GBS) channels (quantum) or an ESN (classical)—combines the encoded current observation with its internal memory to predict the next five-dimensional pressure state. The decoder then uses a rank-32 POD basis and a Gaussian radial basis function (RBF) mapper to reconstruct the full pressure field from the predicted state.

Both dynamical readouts are trained with MSE and L2 regularization on the fitting subset, with hyperparameters selected via validation and refitted on the full training data. Evaluation reports prediction errors on the original pressure scale. For the classical baseline, ESNs with 5--320 nodes are tested over 100 random initializations per size, with results summarized by mean and standard deviation across runs.

For the parameter comparison, we count trainable readout coefficients separately from fixed network coefficients and circuit phase settings, excluding normalization statistics and the common state encoder and spatial decoder.
The photonic model maps 50 measured features to five outputs using $50\times5+5=255$ trainable readout parameters, including output biases.
The 320-node ESN uses $320\times5+5=1{,}605$ trainable readout parameters, corresponding to an $84.1\%$ reduction for the photonic model.
The ESN additionally contains $320\times6+320^2=104{,}320$ fixed input and recurrent coefficients, including input biases, bringing its total to 105,925 network coefficients.
The photonic dynamics are configured by 20 fixed MZI phase settings shared across the five input streams.

\section*{Acknowledgements}
\noindent 
The authors thank Yuxuan Zhang, Gabriele Bressanini, M.S. Kim, and Abolfazl Bayat for helpful discussions.
This research is supported by the National Key R\&D Program of China (Grants No.2024YF
A1409300); National Natural Science Foundation of China (NSFC)(Grant No.62235012, No.12
304342, No.12574549, No.12574542, No.125B1033); Innovation Program for Quantum Science and Technology (Grants No.2021ZD0301500, and No. 2021ZD0300700); Science and Technology Commission of Shanghai Municipality (STCSM) (Grants No.2019SHZDZX01, No.24ZR1438700, No.24ZR1430700 and No.24LZ1401500); Startup Fund for Young Faculty at SJTU (SFYF at SJTU) (Grants No.24X010502876 and No.24X010500170); Frontier Technologies R\&D Program of Jiangsu (Grant No.SBF20250000094). X.-M.J. acknowledges additional support from a Shanghai talent program and support from Zhiyuan Innovative Research Center of Shanghai Jiao Tong University. H. T. acknowledges an open research fund from State Key Laboratory of Photonics and Communications and additional support from Yangyang Development Fund.
\\

\section*{Author contributions}
\noindent
X.-M.J. conceived the project. X.-M.J. and L.Y. supervised the project. Y.-X.F., Y.-L.N., L.-S.Y. and C.-S.X. simulated the integrated photonic devices. Y.-X.F. designed the layout of the quantum computing chip, with assistance from Y.-L.N., L.-S.Y. and H.-Y.S. C.-Y.Z. fabricated the chip. Y.-F.L., H.Y., J.-H.C., Z.-C.W. and Z.-H.L. packaged the chip. Y.-X.F., K.-M.H. and J.-J.H. designed and simulated the algorithms, with assistance from P.-L.D. and Y.-Z.Z. Y.-X.F., H.-Y.S. and X.-Y.W. performed the experimental measurements, with assistance from Y.-L.N., H.S., D.-H.H., C.-Y.P., Y.-Q.P., B.-J.L., Y.-L.Y., Y.X., W.-H.Z., F.L., L.C., X.-T.F and J.-P.D. Y.-X.F., K.-M.H. and J.-J.H. analyzed the experimental data. Y.-X.F. drafted and revised the manuscript. H.-Y.S., K.-M.H., J.-J.H., Y.-R.Z, C.-S.W and H.T. contributed to the revision of the manuscript.

\section*{Competing interests}
\noindent 
The authors declare no competing interests.
\\

\printbibliography{}

\baselineskip21pt
\clearpage

\end{document}